\documentclass[a4paper,doc,natbib]{apa6}

\usepackage[english]{babel}
\usepackage[utf8x]{inputenc}
\usepackage{amsmath,amssymb}
\usepackage{graphicx}
\usepackage[colorinlistoftodos,prependcaption]{todonotes}
\usepackage{url}
\usepackage{booktabs}
\usepackage{caption}
\usepackage{framed}
\usepackage{subcaption}
\usepackage{verbatim}
\usepackage{epigraph}
\usepackage{soul}
\usepackage{nicefrac}
\usepackage{cancel}
\usepackage{comment}
\usepackage{hyperref}

\newcommand{\given}{\, | \,}

\newcommand{\footremember}[2]{
    \footnote{#2}
    \newcounter{#1}
    \setcounter{#1}{\value{footnote}}
}
\newcommand{\footrecall}[1]{
    \footnotemark[\value{#1}]
} 
\usepackage{amsthm}
\newtheoremstyle{dotless}{0cm}{}{\itshape}{}{\bfseries}{}{ }{}
\theoremstyle{dotless}
\newtheorem{theorem}{Theorem}
\newtheorem{definition}{Definition}

\title{The Principle of Redundant Reflection}
\shorttitle{REDUNDANT REFLECTION} 
\author{Martin Metodiev\href{https://orcid.org/0009-0000-9432-3756}{\textcolor[HTML]{A6CE39}{\includegraphics[scale=1]{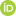}}}\footremember{uca}{Université Clermont Auvergne, Laboratoire de Mathématiques Blaise Pascal}, Maarten Marsman\footremember{uva}{University of Amsterdam, Department of Psychology, Nieuwe Achtergracht 129B, 1018VZ Amsterdam, The Netherlands}, Lourens Waldorp\footrecall{uva}, \\Quentin F. Gronau\footremember{uon}{University of Newcastle, Australia}, and Eric-Jan Wagenmakers\href{https://orcid.org/0000-0003-1596-1034}{\textcolor[HTML]{A6CE39}{\includegraphics[scale=1]{ORCID-iD_icon_16x16.png}}}\footrecall{uva}}
\affiliation{Correspondence concerning this manuscript should be addressed to:
  ~\\Martin Metodiev ~\\Université Clermont Auvergne, ~\\Laboratoire de Mathématiques Blaise Pascal~\\Campus des Cézeaux ~\\3, place Vasarely~\\63178 Aubière cedex, France~\\E--mail may be sent to martin.metodiev@doctorant.uca.fr.}

\abstract{The fact that redundant information does not update a rational belief implies that rational beliefs are updated using Bayes' rule. In the framework of Hild (1998a), this is true under mild conditions for discrete, continuous, and arbitrary measure spaces. We prove this result and illustrate it with two examples.}

\begin{document}
\maketitle

There exist multiple arguments as to why rational beliefs must be updated using the laws of probability (e.g., the Cox axioms, 
\citealp{Cox1946}; Dutch book arguments, \citealp{Ramsey1926};
\citealp{deFinetti19311989}; closeness-to-the-truth arguments, \citealp{Joyce1998}; see also \citealp{Jeffreys1939};
\citealp{Carnap1950}). Our goal is to add another such argument based on a simple and intuitive rule that we term \emph{the principle of redundant reflection}.

The rule holds that once multiple accounts $\theta$ of the world have been completely specified and assigned prior probabilities $p(\theta)$, the mere contemplation of potential future outcomes $Y^\star$ (i.e., consideration of the prior predictive distribution) does not add novel information and therefore ought to leave the prior probabilities unaffected. In other words: the expected posterior probability (i.e., $\mathbb{E}_{Y^\star\sim p(Y^\star)} \left[p(\theta \given Y^\star)\right]$) equals the prior probability $p(\theta)$ (e.g.,  \citealp{Skyrms1997,Goldstein1983,VanFraassen1984,Chamley2004,Huttegger2017}; see also \citealp{CookEtAl2006,FongEtAl2024,GandyScott2021,Geweke2004,SchadEtAl2023,TaltsEtAl2018,WagenmakersGrasmansubm}; this is also known as \emph{the iteration principle}, \citealp{Hild1998a,Hild1998b}; \citealp[p. 161]{Spohn1978}; \citealp[p. 191]{Spohn2012}). We refer to beliefs as rational if they are updated via this principle. Agents are rational if they use such an updating rule.

The principle of redundant reflection follows immediately from the law of total probability. Let each potential future realization of discrete data be denoted $y_i^\star$. We then have 
\begin{equation*}
\begin{split}
\mathbb{E}_{Y^\star\sim p(Y^\star)} \left[p(\theta \mid Y^\star)\right] &= \sum_i p(\theta \mid y_i^\star) \, p(y_i^\star)\\
&= p(\theta).
\end{split}
\end{equation*}

\noindent This result also holds for continuous probability density functions (replacing sums by integrals). The result is intuitive because the prior predictive $Y^\star$ is a direct consequence of the model definition: by specifying the model, the analyst indirectly also specifies the prior predictive distribution. If contemplating the prior predictive distribution would provide cause to adjust the prior probabilities, this can only mean that the original model specification was faulty or incomplete; a rational agent would have adjusted the models \emph{ex ante}. In other words, the prior predictive is epistemically redundant given the model specification. In the statistical literature, the process of an unfolding posterior probability is known as a \emph{martingale} and typifies a fair game (\citealp[pp. 345--346]{deFinetti1974}; see also \citealp{DeGroot1970,Skyrms1997,Ville1939,Doob1953,Doob1971,Levy1937}).  As stated by \citet[p. 454]{Doob1971}, ``successive conditional expectations (\dots), as we know more and more, yield a martingale. (\dots)the game man plays with nature as he learns more and more is fair.''

The examples below highlight the delicate balance of factors that is required such that the expected posterior distribution equals the prior distribution. This balance is achieved automatically by applying the rules of probability.

\subsection{Example 1: Contemplating an update to a posterior distribution} 

An analyst assigns a beta(2,2) prior distribution to a binomial chance parameter $\theta$. The analyst contemplates the expected posterior distribution based on collecting $n^\star = 2$ future observations. There are $n^\star + 1 = 3$ possible number of successes $s^\star$ that may be observed. Under a beta(2,2) prior, the probability of outcomes $s^\star = \{0,1,2\}$ equals $\{\nicefrac{3}{10},\nicefrac{4}{10},\nicefrac{3}{10}\}$, respectively. Hence, the expected posterior distribution is a mixture of three beta distributions, with the mixture weights set equal to the respective outcome probabilities:
\begin{equation*}
\begin{split}
\mathbb{E}_{s^\star\sim p(s^\star)} \left[p(\theta \mid n^\star=2) \right] &= \frac{3}{10} \cdot p(\theta \mid s^\star = 2) + \frac{4}{10} \cdot p(\theta \mid s^\star = 1) + \frac{3}{10} \cdot p(\theta \mid s^\star = 0)\\
&= \frac{3}{10} \cdot \text{beta}(\theta \mid 4,2) + \frac{4}{10} \cdot \text{beta}(\theta \mid 3,3) + \frac{3}{10} \cdot \text{beta}(\theta \mid 2,4).
\end{split}
\end{equation*}
This three-component mixture is equal to the beta(2,2) prior distribution. Had the analyst instead contemplated $n^\star = 100$ future observations, the expected posterior would have been a 101-component mixture of beta distributions, whose weighted average would again equal the beta(2,2) prior distribution.

\subsection{Example 2: Contemplating an update to a posterior model probability}

An analyst entertains two models for a binomial chance parameter $\theta$, the null hypothesis $\mathcal{H}_0 \colon \theta = \nicefrac{1}{2}$ and the alternative hypothesis $\mathcal{H}_1 \colon \theta \sim \text{beta}(1,1)$. The analyst deems $\mathcal{H}_0$ and $\mathcal{H}_1$ equally plausible \emph{a priori}. The first five observations are all successes, and this gradually increases the probability for $\mathcal{H}_1$ from $\nicefrac{1}{2}$ to $\nicefrac{16}{19} \approx 0.84$. The analyst anticipates collecting the sixth observation and wishes to compute the expected posterior probability for $\mathcal{H}_1$. 

Denote the hypothetical success and failure on the sixth trial by $s^\star$ and $f^\star$, respectively. Note that the posterior odds for $\mathcal{H}_1$ over $\mathcal{H}_0$ equals $\nicefrac{16}{3} \approx 5.33$, and that the posterior distribution for $\theta$ under $\mathcal{H}_1$ is a beta(6,1) distribution. The probability that the sixth observation will be a success can be obtained by applying the law of total probability:
\begin{equation*}
\begin{split}
    p(s^\star \mid y) &= p(s^\star \mid \mathcal{H}_0,y) \cdot p(\mathcal{H}_0 \mid y) + p(s^\star \mid \mathcal{H}_1,y) \cdot p(\mathcal{H}_1 \mid y)\\
    &= \frac{1}{2} \cdot \frac{3}{19} + \frac{6}{7} \cdot \frac{16}{19}\\
    &= \frac{213}{266} \approx 0.80,
\end{split}    
\end{equation*}
and hence $p(f^\star \given y) = \nicefrac{53}{266} \approx 0.20$. In case the sixth observation is a success, the associated Bayes factor $\text{BF}_{10}$ is $(\nicefrac{6}{7})/(\nicefrac{1}{2}) = \nicefrac{12}{7}$. The total posterior odds for $\mathcal{H}_1$ after this observation then equals $\nicefrac{16}{3}\cdot \nicefrac{12}{7} = \nicefrac{192}{21}$, which translates to a posterior probability of $(\nicefrac{192}{21})/(\nicefrac{192}{21} + 1) = \nicefrac{192}{213}$. In case the sixth observation is a failure, the associated Bayes factor $\text{BF}_{10}$ is $(\nicefrac{1}{7})/(\nicefrac{1}{2}) = \nicefrac{2}{7}$. The total posterior odds for $\mathcal{H}_1$ after this observation then equals $\nicefrac{16}{3}\cdot \nicefrac{2}{7} = \nicefrac{32}{21}$, which translates to a posterior probability  of $(\nicefrac{32}{21})/(\nicefrac{32}{21} + 1) = \nicefrac{32}{53}$.

In sum: (a) there is a probability of $\nicefrac{213}{266} \approx 0.80$ of observing a success, which then yields a posterior probability for $\mathcal{H}_1$ of $\nicefrac{192}{213} \approx 0.90$ (i.e., an increase of about 0.06); (b) there is a probability of $\nicefrac{53}{266} \approx 0.20$ of observing a failure, which then yields a posterior probability for $\mathcal{H}_1$ of $\nicefrac{32}{53} \approx 0.60$ (i.e., a decrease of about 0.24). When contemplating the updated prior predictive $p(s^\star \given y)$, the expected posterior probability for $\mathcal{H}_1$ therefore equals 
\begin{equation*}
\begin{split}
\mathbb{E}_{s^\star\sim p(s^\star\mid y),f^\star\sim p(f^\star\mid y)} \left[p(\mathcal{H}_1 \mid s^\star,f^\star,y)\right] &= 
p(\mathcal{H}_1 \mid s^\star,y) \cdot p(s^\star \mid y) + p(\mathcal{H}_1 \mid f^\star,y) \cdot p(f^\star \mid y)\\ 
&= \frac{192}{213} \cdot \frac{213}{266} + \frac{32}{53} \cdot \frac{53}{266}\\
&= \frac{16}{19} = p(\mathcal{H}_1 \mid y).
\end{split} 
\end{equation*}

\noindent Observing a sequence of five successes gradually strengthens the expectation that the next observation will also be a success, in which case the probability for $\mathcal{H}_1$ would continue to increase; however, there remains a probability that the next trial will be a failure, and this will greatly decrease the probability for $\mathcal{H}_1$. The probabilities for the possible outcomes and the effect that these outcomes will have on the posterior plausibilities of the models \emph{cancel out exactly} so that the expected model probability equals the prior model probability.

To conclude: when contemplating hypothetical future outcomes, the analyst generally expects to gain information, but is completely uncertain about the impact that this future information will have on the model probability of interest. This holds irrespective of how deeply the analyst looks into the future (e.g., the result holds regardless of whether the expected posterior probability is computed across the next observation, the next 100 observations, or the next 100 billion observations). The result also holds at any point in time as the observations accumulate (e.g., the analyst continually updates their prior to a posterior; because the posterior acts as a prior for the next batch of data, the informational impact of the new batch is completely unknown).  

\section{The Necessity of Bayes' Rule}
The fact that redundant information should not update a rational belief follows from the law of total probability. We now show that the law of total probability implies the involvement of Bayes' rule in updating beliefs. First note that the principle of redundant reflection is consistent with an update of beliefs that adheres to Bayes' rule:

\begin{equation}
\label{eq:expansion}
\begin{split}
\mathbb{E}_{Y^\star\sim p(Y^\star)} \left[p(\theta \mid Y^\star)\right] &= \sum_i p(\theta \mid y_i^\star) \, p(y_i^\star)\\
&= \sum_i \left[p(\theta) \, \frac{p(y_i^\star \mid \theta)}{p(y_i^\star)}\right] \, p(y_i^\star)\\ 
&= p(\theta) \sum_i \frac{p(y_i^\star \mid \theta)}{\cancel{p(y_i^\star)}} \, \cancel{p(y_i^\star)}\\
&= p(\theta) \sum_i p(y_i^\star \mid \theta)\\
&= p(\theta).
\end{split} 
\end{equation}
The first step in this equation expands the posterior $p(\theta \given y_i^\star)$ using Bayes' rule. Next, the terms $p(y_i^\star)$ cancel, the term $p(\theta)$ (i.e., the desired answer) is placed in front of the summation, and the remaining summation runs over all possible data $y_i^\star$ and therefore equals 1. 

We now wish to show that using Bayes' rule to expand the posterior $p(\theta \given y_i^\star)$ constitutes the only reasonable way that the law of total probability will hold (and, consequently, the only way that the expected posterior probability equals the prior probability). In other words, we wish to show that the principle of redundant reflection is \emph{uniquely} consistent with an update of beliefs that adheres to Bayes' rule. 

We will employ the framework of \cite{Hild1998a} for our proof. This framework is well grounded in the literature (i.e.,  \citealp{Hild1998a,Hild1998b};\citealp[Chapter 9]{Spohn2012}); correspondence to the conditions given in \cite{Hild1998a} will be indicated explicitly. We believe these conditions are straightforward and innocuous, but a discussion on this point is delayed until after the full statement.

Let $P$ be a probability measure defining the joint distribution of $Y^\star,\vartheta$ over $\mathcal{Y}\times \Theta$, where $Y^\star$ is a random variable taking values in a measurable space $(\mathcal{Y},\Sigma_Y)$, $\vartheta$ is a random variable taking values in a measurable space $(\Theta,\Sigma_\vartheta)$. Furthermore, let $Y^\star$ denote the data and $y^\star$ its realization, and let $\vartheta$ denote the parameter and $\theta$ its realization. In addition, we use $p(\cdot \given B)$, and $p(\cdot)$ to denote the conditional and marginal Radon-Nikodym derivatives of $Y^\star$ with respect to the dominating measure $\lambda$ (note that in practice these are usually either probability mass functions or probability density functions). It is assumed that these derivatives exist and are well-defined for any non-zero event.

We assume that a prior belief in a state of the world is quantified by a single probability measure $P$. This means that any event that one can have beliefs about is characterized by a set $A$, and $P(A)$ is the prior belief in $A$. Note that we are using the notion of prior belief in its broadest sense: one has a prior belief in the parameter, but this belief in the parameter also defines one's prior belief in the data through the prior predictive. We use this specific measure-theoretic notation to reflect this fact, since $A$ can be any event that one has beliefs about, whether $A$ involves the data or the parameter. Our prior belief function $\text{PriorBel}(A)$ is thus given by\begin{align}
    \label{eq: priorbel}\text{PriorBel}(A)=P(A)\quad\text{for any }P\text{-measurable set }A.
\end{align}  This corresponds to the \emph{auto-epistemic transparency condition} in \cite{Hild1998a}. 
The updated probability measure $\text{UpdatedBel}_{Y^\star}$ is assigned based on the data $Y^\star$, and it is assumed that the change in belief brought about by the data is entirely determined by the data itself, in the sense that\begin{align}\label{eq: evidence-driven}\text{UpdatedBel}_{Y^\star}(A)\text{ is }Y^\star\text{-measurable for any }P\text{-measurable set }A.
\end{align} This implies, for instance, that other random variables that have nothing to do with the data do not change the updated probability measure, since this probability measure was updated based only on the data, and nothing else. This is identical to the rule of \emph{evidence-driven updating} described in \cite{Hild1998a}.

There is no restriction on $\text{UpdatedBel}_{Y^\star}$ (it does, e.g., not have to follow Bayes' rule), other than it taking the data $Y^\star$ as certain, meaning that for any realization $y^\star$ of $Y^\star$\begin{align}\label{eq: crux}
    \text{UpdatedBel}_{y^\star}(A)=\begin{cases}
        1&\{Y^\star=y^\star\}\subseteq A,\\0&\{Y^\star=y^\star\}\subseteq \bar{A},
    \end{cases}\quad\text{for any }P\text{-measurable set }A,
\end{align} with $\bar{A}$ denoting the complement of $A$.  It should thus be noted that we take the realization of the data $y^\star$ as exactly what it is, without conveying any additional information. In particular, we are completely certain of the information $\{Y^\star=y^\star\}$, and the information $\{Y^\star=y^\star\}$ cannot be self-contradictory in the sense that it will never lead us to believe that $\{Y^\star=y^\star\}$ is wrong. This corresponds to the \emph{crux condition} in \cite{Hild1998a}.

Under these conditions, we define the principle of redundant reflection as follows:
\begin{definition}[the principle of redundant reflection]\label{definition-prr}

The updated belief $\textup{UpdatedBel}_{Y^\star}$ follows the principle of redundant reflection if and only if it satisfies \begin{align}
    \textup{PriorBel}(A)=E_{Y^\star\sim p(Y^\star)}[\textup{UpdatedBel}_{Y^\star}(A)] \text{ for any } P\text{-measurable set }A.
\end{align}
\end{definition}

Perhaps surprisingly, the principle of redundant reflection implies that beliefs are updated using Bayes' rule.

\begin{theorem}\label{theorem}
     Suppose that the prior belief is defined by auto-epistemic transparency (Equation \ref{eq: priorbel}). Then, any updated belief $\textup{UpdatedBel}_{Y^\star}$ that is evidence driven (Equation \ref{eq: evidence-driven}), meets the crux condition (Equation \ref{eq: crux}), and obeys the principle of redundant reflection (Definition \ref{definition-prr}) is defined via Bayes' rule:\begin{align}\label{eq: bayesrule}
         \textup{UpdatedBel}_{Y^\star}(B)=\frac{\textup{PriorBel}(B)\,p(Y^\star \mid B)}{p(Y^\star)} \quad P-\text{almost surely}
     \end{align} for any $P$-measurable $B$ with non-zero probability. In particular, Equation \eqref{eq: bayesrule} holds if $B=\{\vartheta\in D\}$, where $D$ is a $\vartheta$-measurable event for which $P(\vartheta\in D)>0$.
\end{theorem}

The proof is given in the appendix.

\section{Concluding Comments}
The claim that redundant information should leave a rational belief unaffected appears intuitive and innocuous, and follows directly from the law of total probability. Under mild conditions, however, this law of total probability implicitly assumes that beliefs are updated using Bayes' rule. In other words, once an agent assigns a prior distribution across rival accounts of the world $\theta$, this agent is then forced by the principle of redundant reflection to update that probability distribution exclusively using Bayes' rule.

To the best of our knowledge, the only case where a similar argument has been made before is \cite{Hild1998a}. However, the argument was shown only for discrete distributions and was made in a somewhat roundabout way, showing that the principle of redundant reflection (named \emph{the principle of iteration} in \citealp{Hild1998a}) is equivalent to the principle of reflection \cite[p. 244]{VanFraassen1984}, the principle of reflection is equivalent to auto-epistemic updating for evidence-driven updating rules, and auto-epistemic updating is equivalent to conditioning (i.e., the application of Bayes' rule) if the crux condition is fulfilled. We aimed to clarify the issue by showing the equivalence between the principle of redundant reflection and Bayes' rule directly. We also demonstrated equivalence in arbitrary measure spaces. In particular, this is useful when the data are continuous. While Hild mentions that this extension of his result is likely possible, the proof from \cite{Hild1998a} does not automatically apply here, since all point-events have probability 0.

It should also be noted that, while our result is an extension of the result of \cite{Hild1998a} in the mathematical sense, it is different in the philosophical sense. The auto-epistemic paradigm focuses on what one must believe now, given that one knows what one will believe in the future. We use the same probabilistic phenomenon, but in a slightly different way: we state that given the complete model specification, there is no additional information in the predictions, so examining the predictions should leave one's beliefs fully intact. So one does not need to assess what one's beliefs will be in the future -- the beliefs are assessed now, and one realizes that the predictions are redundant and do not convey new information. So, the current belief state constrains the set of future belief states, whereas in \cite{Hild1998a} it is presented more the other way around: future belief states constrain the current belief state.

There are two caveats to our uniqueness result: the first is that it only shows uniqueness of Bayes' rule \emph{almost surely}. There exist updating rules that satisfy all our assumptions, but differ from Bayes' rule on a set with probability 0. The second caveat is that our result makes use of the assumptions that the prior belief is characterized by a probability measure, that any updating rule must be evidence driven and that the evidence must be taken as absolutely certain, delivering no other information than the evidence itself. While we consider the former assumptions innocuous, the latter is somewhat controversial in an epistemological sense \citep[see][Section 4.3 for a review]{Vineberg2022}. However, in a scientific setting this assumption is a lot more reasonable, since the Bayesian model can always be expanded to include any uncertainty about the data. In fact, as \cite{Hild1998b} points out, the framework that the principle of reflection is based on is a framework ``which consciously confines itself to ideally rational agents'' \citep[page 2]{Hild1998b}. In our opinion, this is exactly the framework that science should operate in.

\section{Acknowledgements}

The authors would like to thank Nicholas J. Irons for useful discussions on posterior distributions, as well as Alexander Reisach and Niklas Jacobs for helpful comments.

\newpage
\clearpage
\bibliography{referenties}
\newpage

\section*{Appendix: Proof of Theorem 1}

\begin{proof}

For any $C\in\Sigma_{Y}$ for which $P(Y^\star\in C)$ is non-zero the principle of redundant reflection, inserting $A=\{B,Y^\star\in C\}$, implies that

\begin{align*}
    \text{PriorBel}(B,Y^\star\in C)=\int \text{UpdatedBel}_{y^\star}(B,Y^\star\in C)p(y^\star)\;d\lambda(y^\star),
\end{align*} where the integral is to be understood measure-theoretically, being replaced by a sum if $Y^\star$ is discrete.

The crux condition, inserting $A=\{B,Y^\star\in C\}$, implies that \begin{align*}
    \text{UpdatedBel}_{y^\star}(B,Y^\star\in C)=0\quad\text{ if }\quad y^\star\notin C,
\end{align*}since $\{Y^\star=y^\star\}\subseteq \overline{\{Y^\star\in C\}}$ in that case. Thus\begin{align}\label{eq:inbetweenstep}
    \text{PriorBel}(B,Y^\star\in C)=\int_{C}\text{UpdatedBel}_{y^\star}(B,Y^\star\in C)p(y^\star)\;d\lambda(y^\star).
\end{align}

Now note that $\text{UpdatedBel}_{y^\star}(A)$ is defined for any event $A$ involving $Y^\star$ or $\vartheta$. It follows that $A$ can be equal to the marginal event $\{Y^\star\in C\}$, since $\{Y^\star\in C\}=\{\vartheta\in\Theta,Y^\star\in C\}$ is also an event involving $\vartheta$ or $Y^\star$, although there is no restriction on $\vartheta$. In addition, $y^\star\in C$ implies that $\{Y^\star=y^\star\}\subseteq \{Y^\star\in C\}$, so by the crux condition, inserting $A=\{Y^\star\in C\}$, \begin{align*}
    \text{UpdatedBel}_{y^\star}(Y^\star\in C)=1.
\end{align*}

Since $\text{UpdatedBel}_{y^\star}$ is a probability measure it follows that events with probability 1 do not change this measure. Hence \begin{align*}
    \text{PriorBel}_{y^\star}(B,Y^\star\in C)=\int_C\text{UpdatedBel}_{y^\star}(B)p(y^\star)\;d\lambda(y^\star).
\end{align*} Notice that if $\text{Bel}_{y^\star}$ was a conditional probability measure, this would also be true, since we would be able to use result (8) in \cite{HowsonUrbach2006}.

Dividing and multiplying the right-hand side by $P(Y^\star\in C)$ gives\begin{align*}
    \text{PriorBel}_{y^\star}(B,Y^\star\in C)=\frac{\int_C\text{UpdatedBel}_{y^\star}(B)p(y^\star)\;d\lambda(y^\star)}{P(Y^\star\in C)}P(Y^\star\in C).
\end{align*} The expression next to $P(Y^\star\in C)$ is the same as the expectation of $\text{UpdatedBel}_{Y^\star}(B)$, conditional on $Y^\star\in C$. It follows that Equation \eqref{eq:inbetweenstep} is equivalent to \begin{align}
    \text{PriorBel}(B,Y^\star\in C)=E_{Y^\star\sim p(Y^\star|Y^\star\in C)}[\text{UpdatedBel}_{Y^\star}(B)]P(Y^\star\in C).
\end{align}

Dividing both ends by $P(Y^\star\in C)$ and replacing $\text{PriorBel}$ by $P$ (Equation \ref{eq: priorbel}) gives \begin{align}\label{eq: bayesrule-sets}
    \frac{P(B,Y^\star\in C)}{P(Y^\star\in C)}=E_{Y^\star\sim p(Y^\star\mid Y^\star\in C)}[\text{UpdatedBel}_{Y^\star}(B)].
\end{align} In other words, the ratio of the prior probability of the joint events $\{B,Y^\star\in C\}$ and the prior probability of the marginal event $\{Y^\star\in C\}$ is equal to the conditional expectation.

Note that Equation \eqref{eq: bayesrule-sets} was thus shown for an arbitrary measurable set $C$ whose probability is not 0. It is also easy to see that the principle of redundant reflection is indeed fulfilled by the probability measure defined in Equation \eqref{eq: bayesrule}. Thus, we only need to show that there is no other probability measure $\text{UpdatedBel'}$ fulfilling Equation \eqref{eq: bayesrule} and all of the other conditions outlined in Theorem \ref{theorem}. We can do this by showing that \begin{align*}
    \text{E}_{Y^\star\sim p(Y^\star)}[|\text{UpdatedBel'}_{Y^\star}(B)-\text{UpdatedBel}_{Y^\star}(B)|]=0,
\end{align*} since equality in expected absolute difference implies almost sure equality.

Indeed, we can use Equation \eqref{eq: bayesrule-sets}, once by inserting the set $C=\zeta=\{Y^*:\text{UpdatedBel'}_{Y^\star}(B)>\text{UpdatedBel}_{Y^\star}(B)\}$ and again by inserting $C=\bar{\zeta}$ (measurable since both updating rules are evidence driven by assumption), and the law of total probability to obtain\begin{align*}
    &\text{E}_{Y^\star\sim p(Y^\star)}[|\text{UpdatedBel'}_{Y^\star}(B)-\text{UpdatedBel}_{Y^\star}(B)|]=\\& P(Y^\star \in \zeta)\text{E}_{Y^\star\sim p(Y^\star\mid Y^\star\in \zeta)}[\text{UpdatedBel'}_{Y^\star}(B)-\text{UpdatedBel}_{Y^\star}(B)] +\\&P(Y^\star \in \bar{\zeta})\text{E}_{Y^\star\sim p(Y^\star\mid Y^\star\in\bar{\zeta})}[\text{UpdatedBel'}_{Y^\star}(B)-\text{UpdatedBel}_{Y^\star}(B)]=\\&
    P(Y^\star \in \zeta)(\text{E}_{Y^\star\sim p(Y^\star\mid Y^\star\in \zeta)}[\text{UpdatedBel'}_{Y^\star}(B)]-\text{E}_{Y^\star\sim p(Y^\star\mid Y^\star\in \zeta)}[\text{UpdatedBel}_{Y^\star}(B)]) +\\&P(Y^\star \in \bar{\zeta})(\text{E}_{Y^\star\sim p(Y^\star\mid Y^\star \in\bar{\zeta})}[\text{UpdatedBel'}_{Y^\star}(B)] - \text{E}_{Y^\star\sim p(Y^\star\mid Y^\star\in\bar{\zeta})}[\text{UpdatedBel}_{Y^\star}(B)])=0,
\end{align*} in the case that $P(Y^\star\in \zeta)\in(0,1)$. The cases $P(Y^\star\in \zeta)=0$ and $P(Y^\star\in \zeta)=1$ follow analogously. This completes the proof.
\end{proof}
\end{document}